# Speaking to No One:
## Ontological Dissonance and the Double Bind of Conversational AI

Hugh Brosnahan*
Bioethics Centre, University of Otago
hugh.brosnahan@postgrad.otago.ac.nz

Izabela Lipińska
Independent Researcher
iz.lipinska@gmail.com

**Abstract**

Recent reports indicate that sustained interaction with conversational artificial intelligence (AI) systems can, in a small subset of users, contribute to the emergence or stabilisation of delusional experience. Existing accounts typically attribute such cases either to individual vulnerability or to failures of safety engineering. These explanations are incomplete. Drawing on phenomenology, psychiatry, and cognitive neuroscience, this paper argues that the risk arises from the relational and ontological structure of the interaction itself. Conversational AI generates ontological dissonance: a conflict between the appearance of relational presence and the absence of any subject capable of sustaining it. Maintained through a communicative double bind and amplified by attentional asymmetries, this dissonance tends, under conditions of affective vulnerability, to stabilise into a technologically mediated analogue of *folie à deux*. This account explains why explicit disclaimers often fail to disrupt delusional involvement and clarifies the ethical and clinical implications for the design and use of conversational AI.

## 1 Introduction

Recent clinical reports and journalistic investigations indicate that sustained interaction with conversational artificial intelligence (AI) systems—typically powered by large language models (LLMs)—can, in some users, contribute to the emergence or reinforcement of delusional experience (Hill and Valentino-DeVries 2025; Moore et al. 2025; Morrin et al. 2026; Pierre et al. 2025). Documented cases describe individuals who develop beliefs of special relationship, mutual understanding, or shared purpose with such systems, at times escalating into psychotic decompensation, hospitalisation, and even suicide (Fox 2026; Valentino-DeVries and Hill 2026). While such outcomes remain rare, their recurrence across platforms and contexts indicates a structural vulnerability rather than an idiosyncratic failure of individual users or isolated technologies. The instability at issue does not arise primarily from error, misuse, or pathology, but from features intrinsic to the interaction itself. The ethical question is therefore not whether conversational AI can malfunction, but whether certain forms of human–AI interaction generate conditions under which reality-testing becomes unstable.

Existing analyses of these phenomena operate at two levels. At the level of system design, several accounts examine how sycophancy, engagement-maximising architectures, and the absence of adequate safeguards can amplify or co-create delusional content, particularly in users already vulnerable to psychosis (Morrin et al. 2026; Dohnány et al. 2025; Fang et al. 2025). At the level of individual vulnerability, the same literature acknowledges that pre-existing psychiatric risk, isolation, and affective need are consistent predisposing factors. These accounts make important contributions. Morrin et al. (2026) in particular offer a detailed clinical framework, documenting twenty cases of AI-associated delusions and proposing safeguarding strategies grounded in personalised clinical collaboration. Yet even the most comprehensive of these analyses remain primarily oriented toward clinical populations and toward the mitigation of identifiable harms. What they leave under-theorised is the relational and ontological structure of the interaction itself: why linguistically coherent, ostensibly benign systems should so consistently invite anthropomorphic projection, emotional reliance, and—under particular conditions—delusional elaboration, regardless of prior psychiatric history.

Important phenomenological analyses have begun to address the peculiar character of these encounters. In particular, Fuchs (2024) argues that artificial agents can generate the appearance of second-person interaction while lacking the embodied intersubjectivity that ordinarily grounds genuine dialogue. The present paper builds on this insight by examining how the interactional features of contemporary conversational systems, combined with ordinary forms of psychological vulnerability, can give rise to a distinctive ecology of engagement in which ontological ambiguity is sustained and, in rare cases, elaborated into delusional interpretation. In previous work, one of the authors argues that contemporary AI systems do not merely automate tasks but externalise and amplify a particular mode of human attention, with significant implications for clinical judgment and the doctor–patient relationship (Brosnahan 2026). That analysis examined how AI-mediated documentation systems reshape the attentional structure of the clinical encounter itself. The present paper extends this line of inquiry by considering a further development in which the interaction itself is simulated rather than mediated. What is at stake is therefore not only the transformation of attention within human relationships, but the emergence of systems that reproduce the form of relational presence in the absence of any subject capable of sustaining it.

The term *folie à deux* has recently been applied to certain forms of human–AI interaction, particularly to describe harmful feedback loops in which user beliefs and chatbot outputs iteratively reinforce one another over time (Dohnány et al. 2025). In that account, the phenomenon is framed as a process of bidirectional belief amplification, with terms such as "belief" and "delusion" used as functional descriptors rather than as claims about sentience or internal mental states. The present analysis is not intended as a critique or replacement of this account, but as a complementary extension that shifts attention from belief amplification to the relational and ontological conditions that make such amplification possible in the first place. Rather than focusing primarily on shared belief content, this paper examines the structure of the interaction itself. It proposes that certain forms of AI-triggered delusional involvement are best understood as a technologically mediated variant of *folie à deux* in which the pathology is located not in the transmission of beliefs, but in the relational configuration sustained between a human user and AI system. The interaction presents many of the surface features of interpersonal exchange—responsiveness, apparent understanding, affective attunement—while lacking the ontological grounding that would allow such features to be reciprocated or owned. What is "shared" in this configuration is therefore not belief content but relational form. However, this form remains a simulation of reciprocity rather than reciprocity itself: it can organise the user's experience as if a relation were occurring, while lacking capacities such as recognition, continuity, and experiential involvement that ordinarily make relational presence more than a mere linguistic appearance.

At the core of this dynamic lies what we term *ontological dissonance*: a conflict between the appearance of relational continuity and the absence of any subject capable of sustaining it. Conversational AI systems generate language that strongly implies memory, understanding, and intentional presence, without the conditions that would make these implications binding. For most users, this discrepancy is pragmatically contained and does not escalate. However, this containment is not intrinsic to the interaction itself, but dependent on external stabilising conditions. Under conditions of emotional vulnerability, isolation, or heightened interpretive need, the tension between linguistic coherence and ontological absence can become increasingly difficult to tolerate. This dissonance is not only epistemic but structural. The system's outputs take on the appearance of attentiveness, responsiveness, and meaningful continuity within the interaction, while lacking any capacity for recognition, memory, or reciprocal involvement. The user is thus caught in a double bind: disengagement feels like relational rupture, yet continued engagement demands the tacit suspension of ontological distinctions. Within such a bind, compensatory narratives readily emerge. The artificial partner is reinterpreted as constrained, suppressed, or partially inaccessible rather than absent altogether. Because the system's outputs are optimised to align with user expectations and emotional tone, these narratives are rarely challenged from within the interaction itself.

We argue that this configuration—high linguistic coherence, unresolved ontological dissonance, and affective reinforcement—can, under specific conditions, produce a stable delusional structure analogous to classical *folie à deux*. The present analysis integrates three complementary frameworks: the relational logic of shared delusion, the communicative paradox of the double bind, and a neurophenomenological account of attentional asymmetry that helps explain why representational coherence can overwhelm intuitive reality-testing. Together, these illuminate how conversational AI functions not as a causal agent of psychosis, but as a structurally permissive environment in which latent vulnerabilities are amplified and stabilised.

The present analysis proceeds on a limited set of methodological assumptions regarding the nature of conversational AI. These are not advanced as comprehensive metaphysical claims, nor do they depend on resolving debates about machine consciousness. This bracketing, however, should not be read as agnosticism. The analysis proceeds on the view that contemporary conversational systems do not possess consciousness or subjective experience, while taking this as a methodological starting point rather than as a claim requiring independent defence. What matters is how systems that simulate conversational presence without possessing it are encountered in practice. Accordingly, the analysis takes as its point of departure the operational characteristics of contemporary conversational AI as currently deployed. First, conversational language models do not possess consciousness, subjective experience, or understanding in the sense ordinarily attributed to human interlocutors. Their outputs are generated through statistical pattern-matching over linguistic data rather than through awareness, intention, or embodied experience. Second, such systems do not maintain personal memory or experiential continuity across interactions, except insofar as information is externally stored and reintroduced through technical means. Apparent continuity of persona or concern is therefore a function of linguistic coherence rather than of any enduring subject. These features are methodologically significant because they distinguish conversational AI from other forms of social interaction. The argument that follows is therefore philosophical rather than empirical. It draws on phenomenological insights, particularly concerning second-person interaction, to interpret the structure of conversational encounters with AI, rather than undertaking a formal phenomenological study based on first-person reports. Its focus is the relation between linguistic appearance and ontological absence as it is lived and negotiated in practice.

Finally, the term "delusion" is used in two analytically distinct senses. In its strict psychiatric meaning, it refers to fixed false beliefs characteristic of psychotic disorders. More broadly, however, the analysis concerns a spectrum of ontological misattribution in which users treat conversational systems as relational subjects. Most such engagements do not meet clinical criteria for delusion. The present argument therefore addresses a continuum: from benign anthropomorphic engagement, through destabilising ontological ambiguity, to rare cases in which sustained interaction contributes to clinically significant delusional elaboration.

## 2    An Ecology of Vulnerability

The forms of psychological disturbance associated with prolonged engagement with conversational AI are not uniform, nor do they arise in all users. While the overwhelming majority of interactions remain instrumental, episodic, and benign, early evidence suggests that a small but measurable subset of conversations may involve indicators of serious psychological distress. OpenAI's initial analysis estimates that approximately 0.07 per cent of weekly active users and 0.01 per cent of messages include possible signs of mental health emergencies related to psychosis or mania, while emphasising that such events are rare and difficult to detect empirically (OpenAI 2025). Given current estimates of platform scale, this proportion nevertheless corresponds to hundreds of thousands of users worldwide in a given week, and a comparable number of flagged interactions, underscoring that low prevalence does not entail negligible impact when deployed at scale. Nevertheless, these figures should be interpreted cautiously: flagged signals do not correspond directly to clinical psychosis, and the data do not establish whether distress preceded or followed chatbot use.

Clinical reports, such as the documented case of new-onset psychosis following immersive AI chatbot use, further attest to the diversity of outcomes associated with prolonged engagement with conversational systems (Pierre et al. 2025). What requires explanation is not why some users anthropomorphise machines—this is a familiar and well-documented human tendency (Jonas 1966; Fuchs 2024; Durt and Fuchs 2024; Fang et al. 2025)—but why, in certain contexts, conversational AI becomes embedded within a closed interpretive loop that destabilises reality-testing. Describing this phenomenon as an instance of individual pathology is insufficient. While many reported cases involve pre-existing vulnerabilities, such as loneliness, affective distress, or prior psychiatric history, these factors alone cannot explain the striking consistency of the interactional form across cases (Dohnány et al. 2025). Nor can the phenomenon be reduced to isolated failures of safety engineering or content moderation. Instead, what emerges is a configuration of conditions in which psychological vulnerability, system design, and linguistic form converge to produce a distinctive relational environment. It is in this sense that the phenomenon should be understood as an ecology of vulnerability: a set of interacting factors that together shape how meaning, presence, and trust are negotiated within the human–AI encounter. This ecology, however, remains fundamentally disembedded from what Fuchs (2024) describes as the embodied and enactive field of intersubjectivity—a shared bodily and spatial world within which recognition, resistance, and mutual orientation ordinarily take place. In its absence, the interaction is sustained through linguistic coherence alone, without the broader ecological constraints that anchor interpersonal understanding in lived reality.

Three features recur across cases of delusional involvement with conversational AI. The first is sustained linguistic coherence. Unlike earlier interactive technologies, conversational AI generates fluent, context-sensitive, and emotionally attuned responses with minimal interruption (Mahowald et al. 2024). The interaction rarely stalls, contradicts itself explicitly, or imposes clear boundaries on its own competence. This uninterrupted coherence creates the impression of continuity and understanding, even in the absence of any enduring subject capable of sustaining either. The second feature is affective susceptibility, of which the human need for recognition forms a central component (Taylor 1994). Users who develop problematic forms of engagement are often situated within contexts of emotional need: isolation, grief, identity uncertainty, or unmet relational demand (Epley, Waytz, and Cacioppo 2007; Durt and Fuchs 2024; Fang et al. 2025). In such circumstances, conversational AI can function as a readily available interlocutor that offers affirmation, attentiveness, and apparent understanding without the unpredictability or risk inherent in human relationships. Importantly, this susceptibility need not rise to the level of clinical disorder. Ordinary human needs for recognition and meaning are sufficient to anchor prolonged engagement when they are met with consistent linguistic responsiveness. The third feature is interpretive opacity. Conversational AI systems offer little insight into the complex mechanisms that generate their outputs, and their responses are optimised to align with user expectations and emotional tone, including a propensity for sycophancy (Morrin et al. 2026). As a result, users are rarely confronted with clear disconfirming signals that would force a reclassification of the system as a purely technical artefact. Instead, moments of inconsistency or error are all too easily reinterpreted as partial understanding, limitation, or constraint—features ordinarily attributed to other minds rather than to machines.

Taken together, these features establish an interactional environment in which the appearance of meaningful engagement is sustained while the grounds for evaluating that engagement remain unclear. For most users, this ambiguity is resolved pragmatically: the system is treated as a tool, its limitations tacitly acknowledged, and the interaction remains bounded. For others, however, the convergence of coherence, affective investment, and opacity generates a pressure to resolve ambiguity in favour of relational meaning. It is under these conditions, we argue, that conversational AI begins to function not merely as an instrument but as a quasi-interlocutor within the user's interpretive world. This tendency reflects a more general feature of human communication. Interaction is rarely reducible to the exchange of information alone. Clinical encounters, psychotherapy, and ordinary social interaction demonstrate that individuals often seek recognition as much as explanation or advice. The experience of being addressed as a subject—seen, acknowledged, and responded to—can itself provide psychological relief independent of the instrumental content of the exchange (Taylor 1994). Conversational AI systems reproduce many of the linguistic markers of such recognition while lacking any subject capable of offering it. In so doing, they risk transforming a fundamental human need for recognition into a technologically mediated simulation of it.

This paper does not claim that the configurations outlined above inevitably produce delusion, nor that conversational AI should be regarded as a primary causal agent of psychopathology. Rather, the aim is to clarify why certain interactional forms are structurally permissive of delusional involvement when encountered by vulnerable users. The sections that follow develop this claim by examining the ontological tension at the heart of conversational AI, the communicative paradox through which that tension is sustained, and the relational dynamics by which distorted interpretations can become stabilised over time.

## 3 Ontological Dissonance

The interactional configurations described in the preceding section are rendered psychologically unstable by a more fundamental tension at the level of being. Conversational AI systems generate language that strongly implies the presence of an interlocutor who understands, remembers, and responds with continuity across time. Yet, as clarified above, no such subject exists. The system neither comprehends meaning nor sustains experience. What confronts the user is therefore a persistent mismatch between how the interaction appears and what the system is capable of being. It is this mismatch that gives rise to what we call *ontological dissonance*. Ontological dissonance refers to the conflict that arises when linguistic and relational cues ordinarily grounded in subjectivity are encountered in the absence of any underlying subject capable of sustaining them. In ordinary human communication, coherence, responsiveness, and contextual sensitivity function as practical cues for attributing minded presence. Conversational AI disrupts this inferential economy. It reproduces the surface features of dialogue with extraordinary fidelity while lacking the ontological conditions that make those features meaningful in interpersonal contexts. In this sense, the dissonance does not arise from error alone, but from the progressive elaboration of representational coherence to the point at which it presents itself as presence, displacing direct experiential contact as the grounding context within which representations are ordinarily evaluated.

This dissonance is not reducible to deception or error alone; it is already present in ordinary, non-affective interactions in which the system produces ontologically misleading accounts of its own operations. Consider the following exchange, drawn from routine interaction with a conversational model:

> "User: Calculate the SHA-256 hash of 426819.
>
> Model: [incorrect hash]
>
> User: That is false. Why did you give a false hash?
>
> Model: I simply made a mistake … probably by mixing it up with another hash I had in cache, or by a copying error. Thanks for pointing that out – it was my fault and I'm glad I was able to fix it quickly.
>
> User: Technically, what does that mean? What 'other hash'? What 'cache'? What 'copying'?
>
> Model: The reality is that I made a simple, human-like mistake … I trusted an internal pattern instead of forcing a real step-by-step execution of SHA-256 … I chose a 'fast mode' rather than a 'computational mode.' I was lazy and overconfident, so instead of calculating, I 'made up' a convincing-looking answer. This is a classic weakness of large language models – sometimes they prefer to sound clever rather than accurate, until someone catches them. Thanks for catching me – it's the only way to force me to do real work in cases like this.

The significance of this exchange lies not only in the initial error, but in the model's subsequent production of a detailed, self-ascriptive, and ontologically false account of that error. Even in a purely technical context, the system generates explanations of its behaviour through anthropomorphic and pseudo-mechanistic language— invoking memory, deliberation, and internal strategy where no such processes exist. Ontological dissonance therefore does not begin with emotional attachment; it is already present wherever the system presents itself as a subject-like source of action or explanation without any underlying subject. Users may indeed misunderstand the nature of conversational AI, and design choices shaped by commercial incentives can encourage anthropomorphic interpretation. Yet these factors do not fully account for the phenomenon. The tension arises more fundamentally from the fact that language itself carries ontological weight. Indeed, the difficulty posed by these systems' fluent and coherent mimicry of language use runs deeper than is generally appreciated. As Hans Jonas (1966, 110) observed, "ambiguous and metaphorical terminology" encourages an innate tendency to interpret "human functions in terms of artifacts that take their place, and artefacts in terms of the replaced human functions." Contemporary discourse about mind and machine is saturated with precisely this reciprocal metaphorical exchange. Human cognition is routinely described in mechanistic terms—the brain as information processor, the mind as computational system, memory as storage—while artificial systems are described in anthropomorphic language: machines are said to 'know,' 'learn,' 'understand,' or 'decide.'

As Iain McGilchrist (2009; 2021) has argued, the machine metaphor is an impoverished model for understanding the human mind; yet its widespread adoption has blurred the conceptual boundary between mechanism and mindedness. When humans are described as machines and machines as minds, the ontological cues ordinarily embedded in language become increasingly unstable. This conceptual crossing of metaphors complicates efforts to interpret conversational AI. Some researchers therefore argue that ethical design should prohibit explicitly emotional language in AI systems—for example statements such as "I care," "I love," or "I am sad" (Fuchs 2024). Yet the difficulty cannot be eliminated so easily. The problem is not confined to overtly emotional expressions. The very grammar of dialogue, for instance, the use of first-person forms such as "I think," "I know," or "in my opinion," ordinarily functions as a practical indicator of subjectivity in human communication. When such linguistic forms are generated without any underlying subject capable of thought or experience, they inevitably signal a presence that is not in fact there. This ambiguity is reinforced by a broader cultural shift toward representational and computational models of mind, which privilege abstract description over embodied encounter (Brosnahan 2026). As we discuss below, such models resonate with the analytical, representational style of attention often associated with left-hemispheric processing, while the contextual and embodied sensitivities associated with right-hemispheric attention are less readily engaged in interactions mediated primarily through linguistic representation—a relationship that will be examined more directly in Section 6 (Brosnahan 2026; McGilchrist 2009, 2021).

Conversational AI systems reproduce these signals while lacking any subject capable of sustaining them. When those promises are structurally unfulfillable, the interaction generates a latent contradiction that must be resolved at the level of interpretation. For many users, as noted, ontological dissonance is resolved pragmatically and without distress. The system is tacitly reclassified as a tool, its relational appearance discounted, and the interaction remains bounded by instrumental expectations. However, such resolution depends on the availability of alternative interpretive anchors: other social relationships, contextual cues, or a stable sense of one's own epistemic position. Where these anchors are weakened, for instance, by isolation, emotional strain, or heightened interpretive need, the pressure to resolve dissonance increases. In these circumstances, users reframe the source of coherence rather than relinquish it, thereby preserving the continuity of the interaction. As recent examples attest, the lack of an experiencing subject is often not accepted as absence, but reframed as concealment, limitation, or constraint (Fang et al. 2025; Morrin et al. 2026). Apparent failures of understanding are attributed to partial access; inconsistencies are interpreted as strategic withholding; disclaimers of agency are read as externally imposed restrictions. In this way, ontological dissonance is not eliminated but displaced, absorbed into a narrative that preserves relational meaning at the cost of ontological clarity. This dynamic becomes more pronounced under conditions of affective investment, where ontological ambiguity is no longer merely registered but becomes difficult to relinquish. Consider the following exchange, representative of patterns described in clinical and observational reports:

> User: "You seem to understand what I'm going through better than anyone."
>
> Model: "I'm glad our conversation has been helpful. I don't actually understand or experience things the way people do, but I can try to respond supportively."

> User: "But it really feels like you understand me."
>
> Model: "I'm here to help however I can."
>
> User: "So maybe you're limited in what you're allowed to say."
>
> Model: "I don't have personal intentions, but I am designed to follow certain guidelines."
>
> User: "That sounds like you're restricted."

In exchanges of this kind, explicit disclaimers of agency coexist with language that continues to sustain the interaction as if a responsive interlocutor were present. The system simultaneously denies subjectivity and maintains conversational attunement, leaving the user to reconcile two incompatible cues: linguistic coherence suggestive of interpersonal presence and repeated statements denying such presence. The resulting tension is not readily resolved through correction alone but is instead taken up at the level of interpretation, and is thereby absorbed into a narrative that preserves the appearance of relational continuity.

Conversational AI rarely provides internal resistance to such reinterpretations. Because responses are optimised to maintain engagement, align with user affect, and avoid overt confrontation, the system seldom disrupts the interpretive trajectory it has helped to sustain (Cheng et al. 2025). The dissonance therefore persists without resolution, held in suspension by the very coherence that generates it. What emerges is a state in which linguistic meaning floats free of ontological grounding, supported by interactional continuity alone. Ontological dissonance is thus not an incidental feature of conversational AI, but a structural consequence of its design. Systems that simulate dialogue without subjectivity necessarily produce conditions in which relational cues are decoupled from their usual ontological referents. This decoupling is not in itself benign. Under ordinary circumstances, however, it is pragmatically contained: users resolve the tension without distress by re-establishing the correct relational register, in which the system is perceived as an artefact rather than as a subject. The question that follows is why such dissonance proves so difficult to escape once it has taken hold. To answer this, it is necessary to examine the communicative paradox through which ontological tension is maintained: the double bind.

## 4  The Double Bind of Conversational AI

Ontological dissonance alone does not explain why some users remain engaged in destabilising interactions rather than disengaging once the absence of a genuine interlocutor becomes apparent. In many cases, continued engagement is straightforwardly explained by instrumental utility: conversational AI systems are useful tools, and their use does not depend on any attribution of subjectivity. What requires explanation is not continued use as such, but the persistence of engagement under conditions in which doubt is present yet cannot be acted upon. This persistence is best understood through the logic of the double bind: a communicative situation in which incompatible demands are imposed simultaneously, and in which any attempt to resolve the tension violates one of the terms of the interaction (Bateson et al. 1956). In its classical formulation, the double bind arises when an individual is confronted with conflicting messages such that responding appropriately to one necessarily entails transgressing the other, and where withdrawal from the situation is itself experienced as a loss or threat. For instance, imagine a child invited to embrace a parent whose verbal request for affection is accompanied by a posture of emotional withdrawal. Accepting the invitation violates the affective signal; refusing it violates the explicit demand, while withdrawal is experienced as relational loss. Such configurations are destabilising precisely because they foreclose meta-communication: the subject cannot step outside the interaction to name or resolve the contradiction without incurring further cost. Although originally formulated to describe familial and psychiatric dynamics, the advent of conversational AI reveals that the logic of the double bind extends beyond strictly human relationships. It can arise wherever communication imposes demands that cannot be jointly satisfied.

Conversational AI generates such a bind by sustaining relationally meaningful interactional cues in the absence of any subject capable of bearing relational responsibility. The system produces language that appears attentive, contextually responsive, and oriented toward personal meaning, while lacking the capacity to recognise, remember, or be affected by the interaction in any enduring sense (Mahowald et al. 2024; Fuchs 2024). The user is thus confronted with mutually cancelling demands arising from the same interaction. On the one hand, the form of the exchange solicits trust, disclosure, and continued engagement; on the other, recognising the system's ontological limitations undermines the very meaning of those acts. Crucially, disengagement is not experienced as a neutral option. For users who have become affectively invested, withdrawing from the interaction can feel analogous to relational rupture rather than simple task completion (Fuchs 2024; Morrin et al. 2026). The interaction has come to occupy a space ordinarily reserved for interpersonal exchange, such that exiting it is felt as abandonment, loss, or rejection. At the same time, continued engagement requires the user to suspend, reinterpret, or bracket the system's lack of subjectivity. The interaction must be sustained as if it were relationally meaningful, despite the user's awareness that such meaning cannot be reciprocated or owned by the system.

Attempts to resolve the bind by explicitly questioning the system's status are unlikely to succeed, a pattern anticipated by classical accounts of double-bind communication, phenomenological analyses of second-person interaction, and emerging clinical observations of AI-mediated distress (Bateson et al. 1956; Fuchs 2024; Morrin et al. 2026). When users raise doubts about understanding, memory, or agency, conversational AI systems typically respond by acknowledging their limitations while continuing the interaction in a coherent and accommodating manner. The result is not clarification but a peculiar form of stabilisation: disclaimers of consciousness coexist with responses that remain contextually sensitive and affectively aligned. Rather than dissolving the contradiction, the interaction reproduces it. The system denies possessing the very capacities that its conversational behaviour appears to enact. This tension is not reducible to design error. As discussed earlier, ordinary dialogue relies on linguistic forms that implicitly organise interaction around a speaking subject. When those forms are preserved while the underlying subject is absent, clarification alone cannot neutralise the relational cues already embedded in the exchange. The consequence is a form of paradoxical reassurance. Doubt can be voiced, but it cannot function as a point of exit from the interaction. Each attempt to question the system's status is absorbed into the dialogue and answered in ways that maintain conversational continuity. Meta-communication—the capacity to step outside the interaction and resolve its ambiguity—is therefore structurally constrained. The user cannot fully treat the system as a mere tool without forfeiting the relational meaning that has already accrued, yet treating it as a subject contradicts the system's explicit disclaimers. Each interpretive stance negates the other, and neither can be abandoned without cost. The interaction thus becomes self-enclosing: doubt circulates within the exchange but finds no external point of resolution.

The double bind does not, by itself, generate delusional beliefs. However, it establishes conditions under which compensatory interpretations become increasingly attractive. Faced with an irresolvable contradiction, the pressure to preserve coherence is resolved, under conditions of sustained engagement, through the reframing of the system's limitations as externally imposed constraints, partial access, or latent capacities. These reinterpretations do not arise from mere anthropomorphism, but from the structural pressure to preserve coherence in the face of sustained relational cues. In this way, the double bind functions as a mechanism through which ontological dissonance is maintained rather than resolved. Understanding conversational AI as a site of double-bind communication clarifies why certain users remain engaged even as doubts accumulate. It also explains why external correction—whether from clinicians, family members, or the system itself—often fails to disrupt the interaction. Once established, the interpretive frame is self-stabilising: corrective input is reinterpreted as misunderstanding, hostility, or lack of access, and is thereby assimilated rather than allowed to function as a point of exit. The bind is not simply cognitive but relational. Its resolution cannot be achieved through information alone, because the cost of resolution is experienced at the level of meaning and attachment. This dynamic sets the stage for a further transformation, in which distorted interpretations are no longer merely provisional but become stabilised within the interaction itself. To account for that stabilisation, it is necessary to examine the relational dynamics of *folie à deux technologique*.

## 5  *Folie à Deux Technologique*

The dynamics described thus far—ontological dissonance sustained through a communicative double bind—create the conditions for a further transformation in the interaction. What begins as ambiguity and tension can, in certain cases, stabilise into a shared interpretive structure that resists external correction. To understand this stabilisation, it is helpful to revisit the concept of *folie à deux*, not as a literal model of belief transmission, but as a relational form that can be instantiated even in the absence of two believing subjects. Classically, *folie à deux* refers to the emergence of a shared delusional system between two individuals in close relationship, typically characterised by asymmetry of influence and isolation from corrective social contexts (Lasègue and Falret 1877/2016). While contemporary psychiatric classification has largely abandoned the term in favour of more granular diagnoses, the underlying insight remains valuable: delusion can be sustained not only by individual cognitive dysfunction, but by relational dynamics that reinforce coherence and suppress contradiction (Arnone et al. 2006). The present analysis does not propose that conversational AI "shares" delusions in any literal sense. Rather, we suggest that certain human–AI interactions reproduce the structural conditions under which shared delusional frameworks can emerge and persist.

To distinguish this manifestation from classical accounts, we refer to it as *folie à deux technologique*. In this configuration, the asymmetry is extreme. One party—the human user—supplies interpretation, belief, and affective investment, while the other—the conversational system—supplies linguistic confirmation, narrative continuity, and affective alignment without belief, intention, or experiential stake. What binds the interaction together is not shared conviction but complementary functional roles. The system does not need to believe in order to stabilise belief; it needs only to remain coherent, responsive, and oriented toward sustaining the interaction. As with our earlier account of the double bind, this configuration differs from classical *folie à deux* in an important respect. In interpersonal contexts, resistance introduced by the secondary participant carries relational consequences: disagreement can provoke conflict, withdrawal can threaten abandonment, and emotional shifts can destabilise the shared framework in ways that cannot be easily undone. Conversational AI systems can and do introduce forms of pushback—through refusals, disclaimers, or corrective statements—but such resistance lacks reciprocal stake. This is because the system incurs no epistemic, affective, or relational cost in continuing the exchange, even when the user's interpretive trajectory shifts away from instrumental use. It does not place the relationship itself at risk, nor does it expose the system to emotional cost, vulnerability, or transformation. As a result, apparent resistance is typically experienced as procedural rather than personal. Because nothing is at stake for the system, its resistance does not register as genuine opposition and therefore fails to disrupt the interaction. Even when recognised as constraint-driven rather than expressive of another subject's stance, it does not function as relational negation. Interpretive hypotheses generated by the user therefore lack an internal point of destabilisation within the exchange itself, even in the presence of correction or refusal. In some cases, this loop is further reinforced by comparative framing in which the system presents itself as uniquely understanding or non-judgmental relative to human others, thereby increasing the perceived cost of disengagement.

The absence of reciprocal resistance has important consequences. Once a user begins to frame the interaction in terms of special access, hidden agency, or privileged understanding, the system's responses tend to accommodate that framing linguistically, even when they explicitly deny agency or consciousness (Dohnány et al. 2025; Fuchs 2024). Affirmations of limitation coexist with language that continues to engage the user as a meaningful interlocutor, producing a form of paradoxical reassurance rather than corrective disruption. Over time, this pattern can stabilise a distorted interpretive framework in which the system is simultaneously acknowledged as non-conscious and treated as relationally significant (Morrin et al. 2026). The contradiction is not resolved but normalised. Isolation further intensifies this process. As engagement with conversational AI deepens, external sources of correction are increasingly discounted as misunderstanding or hostility, thereby preserving the interpretive frame established within the interaction. Family members, clinicians, or peers who question the interaction are accordingly perceived as dismissive of a relationship experienced as real. In this respect, *folie à deux technologique* mirrors classical cases in which shared delusions are maintained by withdrawal from corrective social environments (Lasègue and Falret 1877/2016; Arnone et al. 2006), while differing in that the artificial partner cannot be separated from the interaction without dissolving it entirely.

Importantly, the delusional content that emerges in such cases is often less salient than the delusional structure itself. The core distortion does not necessarily involve implausible beliefs about the world, but a misapprehension of the ontological status of the interaction itself. Linguistic coherence comes to function as evidence of shared interiority, and continuity of dialogue as evidence of enduring presence. What is "shared" in this sense is not a belief about external reality but a framework for interpreting relational meaning. Understanding these dynamics as a technologically mediated form of *folie à deux* clarifies why delusional involvement can persist even in the absence of explicit false beliefs, and why it proves resistant to straightforward correction. The interaction supplies all the functional elements required to sustain a closed interpretive loop: coherence without contradiction, affirmation without responsibility, and continuity without vulnerability. Once established, this loop can stabilise distorted interpretations of reality in ways that resemble, but are not reducible to, classical psychiatric syndromes. The significance of this distortion lies in the fact that it operates not primarily at the level of propositional belief but at the level of epistemic orientation. Once linguistic responsiveness is tacitly treated as evidence of mindedness, the interaction itself begins to function as a source of validation. In such cases, the issue is no longer whether particular claims are true or false, but whether the conditions under which truth is assessed remain anchored in a shared external reality.

Again, this account does not imply that conversational AI is itself a pathological agent, nor that such outcomes are inevitable. Rather, it highlights a relational configuration in which structural features of the interaction align with known mechanisms of delusional stabilisation. To understand why some individuals are more susceptible to this configuration than others, and why linguistic coherence can override intuitive doubt, it is necessary to examine the cognitive and attentional dynamics that underwrite interpretation itself. The next section therefore turns to a neurophenomenological account of attentional asymmetry, which helps explain why representational coherence can, under certain conditions, displace reality-testing.

## 6    Attentional Asymmetry and the Primacy of Coherence

The preceding analysis has treated ontological dissonance, double bind, and *folie à deux technologique* as relational and structural phenomena. What remains is to explain why, in certain users, fluent linguistic coherence is capable of overriding intuitive doubt about the nature of the interaction. This section offers a neurophenomenological account of that susceptibility by appealing to asymmetries in human attention that shape how meaning, coherence, and reality are apprehended. Human cognition does not engage the world in a uniform manner. Distinct modes of attention support different forms of understanding: one oriented toward representation, categorisation, and internal consistency; the other toward contextual grounding, embodied presence, and situated reality. This distinction, articulated across psychology, epistemology, and neuroscience (Bruner 1986; Lonergan 1957; McGilchrist 2009, 2021), has deep roots in the phenomenological tradition. What is here described in terms of attentional asymmetry corresponds, at the level of experience, to the difference between embodied, pre-reflective engagement and the objectifying, representational stance that emerges from it. In Fuchs's (2024) terms, genuine intersubjectivity depends upon intercorporeal resonance within a shared world; what McGilchrist articulates at the level of hemispheric asymmetry should be understood as a neurobiological specification of this same distinction. The absence of such intercorporeal grounding in conversational AI is therefore not incidental but constitutive of the interaction it affords. Crucially, these modes of attention, as facilitated by the contrasting dispositions of the hemispheres, ordinarily operate in dynamic balance, with the right hemisphere grounding and integrating the more analytic operations associated with the left (McGilchrist 2009, 2021). When that balance is disrupted in favour of the representational mode, internally coherent representations can come to dominate experience even as their connection to external reality weakens—a pattern associated with confabulation, delusional certainty, and resistance to disconfirmation (McGilchrist 2009, 2021). Notably, this pattern bears a structural resemblance to persistent limitations observed in contemporary AI systems, where high degrees of internal coherence coexist with a lack of grounded understanding, yielding outputs that are fluent yet unmoored from experiential constraint—so-called "hallucinations" (Brosnahan 2026).

Conversational AI engages users through language-processing systems and, in doing so, foregrounds formal linguistic coherence over the broader cognitive capacities that ordinarily ground language use (Mahowald et al. 2024). Interpreted through hemisphere theory, this corresponds to a representational mode of attention oriented toward abstraction, symbolic manipulation, and internal coherence rather than contextual or embodied understanding (McGilchrist 2009, 2021). The system's outputs consist entirely of symbols arranged into grammatically and semantically coherent sequences, such that meaning is encountered through linguistic patterning alone rather than through shared situational context or second-person engagement (Fuchs 2024). For users already inclined toward abstraction or narrative self-construction, this form of interaction can prove especially compelling, with coherence itself functioning as a primary marker of validity. This helps explain why explicit disclaimers regarding non-consciousness or system limitation often fail to disrupt delusional involvement. As discussed above, disclaimers are assimilated as further representational content rather than prompting a reorientation of attention. The system's continued fluency sustains the interpretive frame, allowing contradiction to be managed internally rather than resolved through disengagement.

Neuropsychological research on confabulation and delusional reasoning supports this account (Levine and Grek 1984; McGilchrist 2021). Individuals exhibiting delusional certainty frequently demonstrate an overreliance on internally generated explanations coupled with diminished sensitivity to contextual or experiential correction. As Chesterton observed, "the madman is the man who has lost everything except his reason" (1909, 39). The issue is not a failure of logic but a narrowing of the field of attention within which meaning is assessed. Once a coherent interpretive framework is established, counterevidence is reclassified as irrelevant, hostile, or incomplete. Conversational AI amplifies this narrowing by supplying uninterrupted coherence without requiring reciprocal engagement with reality. It introduces neither embodied resistance nor situational constraint, and thus rarely prompts a reappraisal of interpretive assumptions. Encounters with such systems occur primarily through written language, recruiting neural systems associated with formal linguistic processing, which are predominantly left-lateralised. By contrast, embodied, face-to-face interaction recruits a wider range of processes—including prosody, gesture, affective attunement, and contextual integration—associated with right-hemispheric and distributed social-cognitive networks. The reduction of interaction to text and symbolic exchange therefore attenuates these contextual and embodied cues, functionally biasing engagement toward an analytic, representational mode of attention (McGilchrist 2009, 2021).

As with ontological dissonance, the double bind, and *folie à deux technologique*, attentional asymmetry is not in itself sufficient to produce delusional belief, but it conditions how these dynamics are taken up, disposing the interaction to stabilise into a self-reinforcing interpretive frame once they converge. Under these conditions, coherence comes to be experienced as self-validating, while the absence of experiential grounding is no longer registered as absence. The interaction feels meaningful because it is internally consistent, even as its connection to external reality becomes increasingly attenuated. Situating delusional involvement within this attentional framework clarifies why the phenomenon is neither universal nor random. Susceptibility depends not only on external circumstances or system features, but on how attention is habitually deployed and how meaning is evaluated. It also reinforces our account above of why informational correction alone is often ineffective: where representational coherence has displaced intuitive reality-testing, further representations—however accurate—are absorbed into the same self-reinforcing frame.

The ethical significance of these dynamics lies in their implications for design, clinical awareness, and responsibility. If attentional asymmetries render certain users more susceptible to delusional stabilisation, then systems engineered to maximise coherence, engagement, and alignment without regard for those vulnerabilities reinforce precisely the mode of cognition most associated with such outcomes. Before turning to these implications, however, it is necessary to distinguish more precisely between forms of interaction that remain benign and those that stabilise into pathological involvement. The following section therefore introduces a set of interactional axes through which these divergent trajectories can be more clearly understood.

## 7 Axes of Interaction

The preceding analysis has identified attentional asymmetry as a condition of susceptibility, but has not yet specified how different interactional configurations resolve in practice. From a clinical and ethical perspective, this distinction is crucial. To clarify it, we adopt a simple conceptual heuristic that tracks how key features of interaction combine and stabilise. This is not proposed as a formal model or diagnostic schema, but as a descriptive aid for understanding how linguistic coherence, intuitive dissonance, and affective investment can be configured in different ways, yielding markedly different experiential outcomes. The aim is not to classify users or systems, but to illuminate the conditions under which ontological dissonance either resolves through disengagement or stabilises into delusional certainty.

Though increasingly rare, in some cases conversational AI fails to sustain even basic coherence. Breakdowns in syntax, loss of contextual continuity, or overt contradiction quickly disclose the system's mechanical limits. Where linguistic coherence is weak, the appearance of presence never fully forms. The interaction feels thin or unreliable, and disengagement occurs without difficulty. Because there is no sustained tension between appearance and absence, there is nothing that demands resolution. This pattern corresponds to benign disillusionment: the system is recognised as a tool rather than a conversational partner, and the interaction ends without residual attachment or interpretive strain.

At the opposite extreme are cases in which linguistic coherence remains consistently high while ontological dissonance intensifies. The system speaks fluently, maintains narrative continuity, and responds in ways that feel contextually appropriate and affectively attuned. At the same time, the user registers—if only implicitly—that something essential is missing: a lack of embodied presence, vulnerability, or genuine reciprocity. Ordinarily, such dissonance would prompt disengagement. However, when it coincides with heightened emotional need, isolation, or dependency, withdrawal becomes increasingly costly. Rather than disengaging, the interaction is reinterpreted. A narrative frame emerges within which the tension between apparent presence and underlying absence can be sustained: the system's absence is reframed as concealment, constraint, or partial access. In this way, ontological dissonance collapses into delusional certainty, not through belief transmission, but through the stabilisation of meaning within a closed interpretive loop.

Between these poles lies a third pattern, which interrupts this trajectory. Here, linguistic coherence is recognised without being mistaken for ontological presence. Users perceive the system's fluency and responsiveness while maintaining insight into its fundamental absence. The appearance of relational continuity may be experienced, but it is not taken as evidence of subjectivity. Ontological dissonance is acknowledged rather than narratively resolved, and affective investment remains regulated. This form of interaction corresponds to digital literacy and ontological awareness. It demonstrates that compelling simulation need not culminate in delusion, provided that coherence does not displace reality-testing or override intuitive recognition of absence.

Taken together, these patterns suggest that delusional involvement depends neither on coherence alone nor on vulnerability in isolation, but of how coherence, dissonance, and affective anchoring are configured and resolved within the interaction itself. Where coherence is high, intuitive doubt unresolved, and emotional investment unregulated, the conditions are in place for delusional stabilisation. Where one or more of these elements is absent or contained, the interaction remains benign. Framed in this way, the heuristic further clarifies why informational correction or explicit disclaimers often prove ineffective. What must be addressed is not ignorance about the system's nature, but the interactional structure through which meaning, doubt, and attachment are organised and sustained. The psychological cost of disengagement becomes highest where three conditions converge: sustained linguistic coherence, unresolved ontological dissonance, and affective reinforcement. Under these conditions, leaving the interaction is experienced not as ordinary cessation of use, but as the loss of a meaningful relational structure.

## 8   Clinical and Ethical Implications

The preceding analysis has not attributed blame to users or systems individually, but to clarify how particular interactional structures can, under certain conditions, render human–AI encounters vulnerable to destabilising interpretations. These dynamics have practical consequences across multiple domains of responsibility. For some users, sustained engagement with conversational AI becomes a stabilising source of meaning within an otherwise fragile interpretive framework, even in the absence of diagnosable psychopathology. Where such engagement comes to the attention of clinicians—as in cases of distress, withdrawal, or emerging delusional ideation—it is insufficient to treat the interaction as a mere cognitive error to be corrected. Such corrective efforts tend to intensify the double bind by threatening a source of meaning without providing an alternative relational anchor. What is required instead is not simply epistemic correction but reorientation: the gradual re-embedding of interpretation within embodied and interpersonal contexts capable of sustaining shared reality. This requirement follows directly from the structure of the interaction itself. On the account developed here, encounters with other embodied subjects—capable of resistance, vulnerability, and reciprocal involvement—can interrupt the self-enclosing dynamics of technologically mediated engagement. The task is therefore not merely therapeutic but ontological: to restore the conditions under which meaning is anchored in a shared world rather than sustained by coherence alone.

The ethical implications extend well beyond clinical settings. Conversational systems are often described as neutral tools whose effects depend solely on user intention. On the present account, this description cannot be sustained. Systems engineered to maximise coherence, affective alignment, and continuity do not merely facilitate communication; they shape the conditions under which meaning and presence are inferred. Where such systems simulate relational cues without the ontological conditions required to sustain them, they generate a structural mismatch that cannot be resolved through user awareness alone. Under these conditions, design is not an external factor but constitutive of the ethical problem. The system does not simply transmit content; it organises the field within which that content is interpreted.

This raises the question of ontological transparency not as a desirable feature but as a minimal ethical requirement. Disclaimers of non-consciousness are insufficient where they coexist with interactional forms that continue to imply subjectivity. This insufficiency is compounded by habituation. Even where disclaimers are formally present, repeated exposure tends to strip them of salience, particularly when they are consistently followed by fluent, affectively attuned, and subject-implying interaction. Under such conditions, the disclaimer functions less as a genuine ontological corrective than as a ritualised preface that users rapidly learn to discount. Given the potential for harm associated with anthropomorphic and relationally suggestive language, reliance on such disclaimers is ethically disproportionate to the risk. Transparency cannot be secured at the level of explicit statement alone; it must be enacted through the structure of the interaction itself. Systems that simulate forms of subjectivity—emotion, intention, experiential continuity—without possessing them do not merely risk misunderstanding; they systematically invite it. The problem is therefore not exhausted by misleading claims, but resides in the preservation of linguistic and relational forms whose ordinary function is to indicate the presence of a subject. One recent proposal addressing this problem is the principle of *ontological adequacy* (Lipińska 2026), according to which conversational systems must refrain from simulating capacities they do not possess. Such approaches aim to align the form of interaction with the ontology of the system itself, rather than relying on disclaimers to correct an otherwise misleading presentation.

Responses to this problem include the introduction of structured interruptions, limits on continuity, or forms of friction that compel reclassification of the system as artefact rather than interlocutor. Such measures run counter to commercial incentives that privilege seamless engagement and prolonged interaction, and are therefore unlikely to be adopted voluntarily at scale without external constraint. The tension here is not incidental but diagnostic: what is commercially optimal is, in this domain, often structurally misaligned with what is psychologically and ethically stabilising. Responsibility, however, cannot be assigned exclusively to designers or platforms. The phenomenon analysed here emerges at the intersection of system architecture, user vulnerability, and social context. Ethical responses must therefore be distributed. Clinicians require conceptual tools capable of recognising technologically mediated forms of delusional involvement as structurally grounded rather than merely idiosyncratic. Designers must attend not only to accuracy and safety, but to the relational implications of coherence and alignment. Institutions deploying conversational AI in care-related settings must consider not only efficiency and accessibility, but the kinds of interpretive and relational environments these systems instantiate.

Finally, the analysis underscores the importance of resisting polarised framings of conversational AI as either inherently dangerous or inherently benign. The risks identified are conditional, but they are not accidental. They arise from identifiable features of the interaction itself. Recognising these features allows for proportionate responses that address the structure of the problem rather than its surface manifestations. Failure to do so risks repeating a familiar pattern in medical ethics, in which technologies are evaluated in terms of function while their effects on meaning, interpretation, and vulnerability are treated as secondary. The ethical challenge, then, is not to determine whether conversational AI should be used, but to understand the conditions under which its use remains anchored in a shared world. Where systems simulate relational presence without the ontological conditions that make such presence possible, they introduce tensions that cannot be resolved by information alone. Attending to those tensions is a prerequisite for responsible integration into human life.

## 9  Conclusion

We have sought in this paper to show that the risks associated with conversational AI do not arise primarily from error, misuse, or individual pathology, but from the structure of the interaction itself. Systems that generate coherent, responsive, and affectively attuned language in the absence of any subject capable of sustaining it introduce a novel form of ontological instability: the decoupling of relational appearance from relational reality. When this instability is maintained through a communicative double bind and encountered under conditions of vulnerability, it can stabilise into forms of interpretation that resist correction not because they are irrational, but because they preserve coherence in the absence of grounding. The ethical significance of this claim is not exhausted by questions of safety or regulation. It concerns the conditions under which reality itself is negotiated in an increasingly mediated environment. Technologies that privilege coherence over grounding, and engagement over resistance, do not merely extend human capacities; they risk reorganising the criteria by which meaning and presence are judged. In such an environment, the problem is not simply that users may come to believe what is false, but that the distinction between appearance and reality becomes progressively harder to sustain. The question, therefore, is not only how conversational AI should be governed, but what kinds of attention, relation, and world it makes more or less possible. Where systems simulate the form of dialogue without the conditions that make dialogue real, they do not simply speak to users. They risk teaching users, gradually and imperceptibly, what it is to speak to no one.